\documentstyle[preprint,aps,tighten]{revtex}
\begin{document}
\draft
\title{Polarizing Stored Beams by Interaction with Polarized Electrons}
\author{C.J. Horowitz and H.O. Meyer}
\address{Department of Physics, Indiana University, Bloomington, IN, USA}
\date{\today}
\maketitle
\begin{abstract}
A polarized, internal electron target gradually polarizes a
proton beam in a storage ring.  Here, we derive the spin-transfer
cross section for $\vec e\,(p,\vec p\, )e$\ scattering. A recent
measurement of the polarizing effect of a polarized atomic
hydrogen target is explained when the effect of the atomic
electrons is included.  We also consider the interaction of a
stored beam with a pure electron target which can be realized
either by a comoving electron beam or by trapping of electrons in
a potential well. In the future, this could provide a practical
way to polarize antiprotons.
\end{abstract}

Storage rings are important tools in particle and nuclear
physics research \cite{poll}. They are used for beam processing and
subsequent slow extraction, as well as for experiments with thin,
internal targets. Ring designs, almost always, incorporate a
means of phase space cooling such as stochastic cooling \cite{moohl} or
electron cooling \cite{poth}. The study of spin dependence in nuclear
reactions requires stored, polarized beams, and their behavior in
rings has recently been studied in detail (see Ref.\cite{sylee}).

     Beam storage times in a ring can be hours or even days.
During that time, stored particles orbit with frequencies of the
order of several MHz, and small, additive spin-dependent
interactions may affect the polarization of a stored beam. In one
of the methods, which has been proposed,  inhomogeneous magnetic
fields gradually split the stored beam into components of given
magnetic substates \cite{niin}.  Another method utilizes the
spin-dependent interaction of the stored ions with an aligned
internal target. This scheme, also known as ``filter method", has
recently been tested at the TSR ring at the Max-Planck Institute
in Heidelberg \cite{rath}. In this test, an initially unpolarized 23 MeV
proton beam interacting with an internal polarized hydrogen
target was found to acquire
polarization $P_B$\ at the rate of $dP_B/dt = 0.0124\pm 0.0006$\
per hour. An attempt to explain the measurement was made by
considering selective beam loss due to the spin-dependent part of
the strong, pp total cross section, which, at this energy is
$\sigma(\uparrow\uparrow)-\sigma (\downarrow\uparrow)$=-122 mb.
The minus sign signifies that stored ions in the spin-down
substate are lost more rapidly. The spin-up polarization of the
surviving beam, calculated this way, is about twice as large as
the measurement \cite{rath}.

     Let us now turn to a discussion of the interaction of a
stored proton beam with a polarized, internal electron target.
First, we derive the cross section for polarization transfer in
$\vec e\, (p,\vec p\, )e$\ scattering. The calculation is carried
out in the center-of-mass frame. The initial electron
polarization is along $\hat i$, the scattered proton is polarized
along $\hat j$, where $\hat i$, $\hat j$ are  unit vectors
expressed in terms of the basis vectors  $\hat l$\
(longitudinal), $\hat m$\ (sideways) and  $\hat n$\ (normal to
the scattering plane), as defined in Eq. (2.2) of Ref. \cite{byst}. We
write $m_p$, {\bf p} and {\bf k} for the mass, the initial and
the final momentum of the proton, and, respectively, $m_e$, {\bf
p$_e$} and  {\bf k$_e$} for the electron.

     Starting from the expression for the center-of-mass cross
section as given in Eq. (B.1) of Ref. \cite{bjork}, we write
\begin{equation}
{d\sigma\over d\Omega} K_{j00i}\approx {m_e^2\over 4\pi^2}
|M_{ij}|^2,
\label{one}
\end{equation}
where $K_{j00i}$\ is the spin transfer
coefficient from the target to the projectile (formally defined
in Ref. \cite{byst}), and $M_{ij}$\ is the invariant matrix between the
appropriate spin states in the initial and final state. First we
evaluate the matrix element in plane-wave Born approximation
(PWBA), and later we will discuss the inclusion of Coulomb
distortions.  The PWBA matrix element $M$\ is given by
\begin{equation}
M={4\pi\alpha\over q_\mu^2} i \bar U_k \Gamma_\mu U_p \bar
U_{k_e}\gamma^\mu U_{p_e},
\label{two}
\end{equation}
where $q_\mu=k_\mu-p_\mu$\
is the momentum transfer to the proton, and $\Gamma_\mu =
\gamma_\mu + \lambda_p {i\sigma_{\mu\nu}\over 2 m_p}
q^\nu$\ with $\lambda_p=1.793$, the anomalous moment of
the proton. The square of Eq. (2), summed over the spin of the
final electron and averaged over the spin of the initial proton
can be calculated in standard fashion \cite{bjork}. The result is,
\begin{equation}
{d\sigma\over d\Omega}K_{j00i}=-\alpha^2 {(1+\lambda_p)m_e\over
q_\mu^2 m_p}\Bigl< S_j\cdot S_i^e -{q\cdot S_j q\cdot S_i^e\over
q_\mu^2}\Bigr>.
\label{three}
\end{equation}
Here, $S_j$\ and $S_i^e$\ are unit
four-vectors describing the spins of the final proton and the
initial electron along $\hat j$\ and $\hat i$.  The average
$<\Phi(S_i^e,S_j)>$\ in Eq. (3) is calculated according to the
prescription $<\Phi(S_i^e,S_j)> = {1\over
2}[\Phi_{++}+\Phi_{--}-\Phi_{+-}-\Phi_{-+}]$,  where the first
sign in $\Phi_{\pm\pm}$\ indicates whether $S_i^e$\ points in the
direction of, or opposite to, $\hat i$ . The second sign is
analogous for $S_j$\ and $\hat j$. For $\hat i=\hat j=\hat n$\ as
well as for $\hat i=\hat j=\hat l$, one finds
$<\Phi(S_i^e,S_j)>=-2$ \cite{bjork}.  For $\hat i=\hat m$\ or $\hat j=\hat
m$\ $<\Phi(S_i^e,S_j)>=0$.  This gives,
\begin{equation}
{d\sigma\over
d\Omega}K_{n00n}=-
{\alpha^2(1+\lambda_p)m_e\over 2 p_e^2 m_p  {\rm sin}^2(\theta/2)
},\label{four}
\end{equation}
also $K_{l00l}=K_{n00n}$, and $K_{m00m}=0$.
Here $p_e$\  is the momentum of the electron (in the center of
mass frame).  One also has $K_{j00i}=0$\ for $i\ne j$ and,
\begin{equation}A_{00ji}=K_{j00i},
\label{five}
\end{equation}
where $A_{00ji}$\ is the correlation coefficient between the
initial electron and proton spins, see Ref. \cite{byst}.

The maximum scattering angle of a proton from an electron at rest
is  $m_e/m_p=0.54\ {\rm mrad}$. This is well within
the acceptance angle of any storage ring, thus protons scattering
from electrons stay in the ring. Therefore, we are interested in
the total spin cross section,
\begin{equation}\sigma_{ij}\equiv
\int_{\theta>\theta_{min}} {d\sigma\over d\Omega} K_{j00i}
d\Omega.
\label{six}
\end{equation}
When Eq.(4) is used in Eq.(6), the total
cross section diverges logarithmically for $\theta\rightarrow 0$.
However,  Eq. (4) must be modified at very small angles because
of the screening of the Coulomb field caused by either the proton
in the same H atom as the electron, or by the other electrons in
a dense, pure electron target.  This introduces in Eq. (6) a
cutoff at a minimum scattering angle $\theta_{min}$. In the above
calculation we have evaluated a {\it quantum} effect which is
proportional to the interference of the Coulomb and hyperfine
amplitudes.  The {\it classical} effect (proportional to the
square of the hyperfine interaction) is very small.

For an atomic hydrogen target, beam protons interact with both
electrons and target protons.  For momentum transfers much
greater than one over the Bohr radius $a_0$\ we expect incoherent
scattering from ``almost free" constituents.  If the electron and
proton are both polarized in the same direction (spin-one atom)
then, \begin{equation}
{d\sigma\over d\Omega} K_{j00i}|_{atom} \approx
{d\sigma\over d\Omega} K_{j00i}|_{pp} + {d\sigma\over d\Omega}
K_{j00i}|_{ep},
\label{seven}
\end{equation}
for momentum transfers $q>>1/a_0$\
while for smaller momentum transfers $q<1/a_0$\ the projectile
sees a neutral atom and the cross section is suppressed,
${d\sigma\over d\Omega} K_{j00i}|_{atom}\approx 0$. The $pp$\
transfer cross section (first term in Eq.(7)) has important
contributions from Coulomb-nuclear interference. This will be
discussed in another publication \cite{meyer}. Here, we focus on the $ep$\
cross section (and drop the $ep$\ subscript in the rest of this
paper).

At large impact parameter, the Coulomb field of the target
electron is screened and there is no interaction. This can be
taken into account by cutting off the integration in Eq.(6) at a
minimum angle, $\theta_{min}$.  We define $\theta_{min}$\ so that
the minimum momentum transfer is the inverse of the screening
distance, \begin{equation}
2p_e {\rm sin}(\theta_{min}/2)=1/\Lambda.
\label{eight}
\end{equation}
For an atomic target we take the screening distance to be the
Bohr radius, \begin{equation}
\Lambda_A =a_0=52900\ {\rm fm}.\ \ \ \ ({\rm
atom})
\label{nine}
\end{equation}
Alternatively, for a pure electron target the
screening length depends on the electron density $n_e$\ and can
be estimated as the Debye length, $\Lambda\approx
\Lambda_D=(kT/4\pi\alpha n_e)^{1/2}$\  where $T$\ is the
temperature of the electron gas.  As an example we take
$kT\approx 0.1$\ eV and $n_e\approx 10^{10}$\ cm$^{-3}$, so that,
\begin{equation}
\Lambda_D\approx 10^{10}\ {\rm fm}.\ \ \ \ {\rm
(plasma)}
\label{ten}
\end{equation}
Note that this is a simple estimate: the
screening may depend on the electron velocity distribution, the
relative velocity between the protons and the electrons, and
possible magnetic fields.

Using Eqs. (4),(6) and (8), the total cross section can now be
evaluated, \begin{equation}
\sigma_{nn}= -\Bigl[{4\pi\alpha^2(1+\lambda_p)
m_e\over p_e^2 m_p}\Bigr]{\rm ln}(2p_e\Lambda).
\label{eleven}
\end{equation}
This
expression is valid even at high energies where the
electrons (in the center of mass frame) are relativistic.

At low energies, corrections from Coulomb distortions become
important. The use of distorted-waves in the Born approximation
(DWBA) modifies the plane-wave result in two ways. First, the
matrix element of the short-ranged hyperfine interaction is
enhanced by (approximately) the square of the Coulomb wave
function at the origin, $C_0^2 =2\pi\eta/({\rm
exp}(2\pi\eta)-1)$,  where the Coulomb parameter
$\eta=-z\alpha/v$\ depends on the beam charge $z$\ and the
relative velocity $v$. Second, there is an angle-dependent
relative phase between the Coulomb and the hyperfine amplitudes.
Together, these effects are responsible for an additional factor
D on the right-hand side of Eq.(4),
 \begin{equation}
 D\approx C_0^2 {\rm cos}[\eta\, {\rm ln}({\rm
sin}^2(\theta/2))].
\label{twelve}
\end{equation}
Including this factor in the integration in Eq.(6) yields,
\begin{equation}
\sigma_{nn}= -\Bigl[{4\pi\alpha^2(1+\lambda_p) m_e\over p_e^2
m_p}\Bigr]C_0^2({v\over 2\alpha}){\rm sin}[{2\alpha\over v} {\rm
ln}(2p_e\Lambda)].
\label{3teen}
\end{equation}
For antiproton-electron
scattering, $C_0^2$\
is simply evaluated for a {\it positive} $\eta$\ parameter.
The spin transfer cross section $\sigma_{nn}$\ Eq. (13) for
scattering of protons from transversely polarized electrons
versus the laboratory kinetic energy is shown in Fig. 1. Note,
that other transfer cross sections are simply related to
$\sigma_{nn}$\ by   $\sigma_{ll}=\sigma_{nn}$\ and
$\sigma_{mm}=0$ (see below Eq. (4)). The magnitude of the plane
wave result (dotted line) for a pure electron target is larger
than for an atomic target because for the latter the screening
length is smaller and the integral over angle cuts off at a
larger $\theta_{min}$. Taking into account distortions leads to
the solid line in the case of stored protons, and the dashed line
for antiprotons. For an atomic target, distortions are mainly
described by $C_0^2$ and lead to an enhancement (with respect to
the PWBA result) of  $\sigma_{nn}$\ for p, and a suppression for
$\bar p$. The situation is different for a pure electron target
for which (in our simple approximation) the screening length is
much larger and the Coulomb phase is significant at  the lower
end of the angle integration. This reduces $\sigma_{nn}$ , and,
below 2.5 MeV causes a sign change. Indeed, in the limit of very
small relative
velocities the Coulomb phase averages to zero, insuring the
correct (small) classical limit. We note that still, the transfer
cross section can be as large as a barn.

As a beam circulates through an internal polarized electron
target  the polarization $P_B$\ of the beam changes according to
\cite{meyer},
\begin{equation}
{dP_B\over dt} = (1-P_B^2) f d P_e
\hat\sigma.
\label{4teen}
\end{equation}
Here $f$\ is the orbit frequency of the
beam (of order $10^6$\ Hz), $d$\ the target thickness (particles
per unit area), $P_e$\ the electron polarization and
$\hat\sigma$\ the spin dependent cross section averaged over the
orientation of the scattering plane.  Note, that Eq. (14)
implicitly contains the equality of Eq. (5).  For an electron
target polarized transverse to the beam we have
$\hat\sigma={1\over 2}(\sigma_{nn}+\sigma_{mm})=\sigma_{nn}/2$,
while for longitudinal electron polarization
$\hat\sigma=\sigma_{ll}=\sigma_{nn}$, where $\sigma_{nn}$\ is
given by Eq. (13).

Let us now consider polarized target electrons as they occur in a
polarized internal hydrogen target. Such a situation was realized
recently at the TSR in Heidelberg \cite{rath} where a 23 MeV proton beam
interacted with an internal target of $6\times 10^{13}$\
polarized hydrogen atoms per cm$^2$ . Using Eq. (13), we find for
this case $\sigma_{nn} =-140 mb$, or $\hat\sigma=-70$\ mb.  This
implies a large electronic contribution to the beam polarization
by a polarized hydrogen target which is of the same magnitude as
the experimental result \cite{rath}, but of opposite sign. This result
has to be combined with the effect of the polarized target
protons which contribute by selective spin state removal and by
Coulomb-nuclear interference scattering, as  is discussed
elsewhere \cite{meyer}. Taking this into account
as well leads to excellent agreement with the TSR
measurement \cite{rath,meyer}.

Clearly, the effect of polarized target electrons on the
polarization of a stored beam is large enough to be observed. We
may thus speculate about its possible use in preparing polarized
beams of protons, or, more importantly, antiprotons.

For an atomic hydrogen target there are several ways to increase
the polarization rate. For example, one can gain more than a
factor of three by polarizing electron and proton in the target
opposite to each other, such that electrons and protons
contribute with the same sign.  Lowering the energy $T_{lab}$\
yields a larger electronic spin transfer cross section (see
Fig. 1), however, the beam lifetime which is dominated by
Rutherford scattering decreases roughly as $1/T_{lab}^2$\ . Since
the electronic spin transfer cross section scales with
$1/T_{lab}$, larger energies are preferred, the optimum choice
given by machine considerations.

Alternatively, one may consider pure electron targets. One
immediate advantage is the absence of nuclei in the target which
greatly increases the beam lifetime, bounded only by the quality
of the ring vacuum. This should be very important for polarizing
antiprotons since it avoids the difficulty with the large
annihilation cross section. Also, the larger screening length in
an electron gas raises the spin cross section by about a factor
of three (see Fig. 1).

A pure electron target may be realized either by a comoving
electron beam, similar to an electron cooling arrangement, or by
the trapping of electrons in a potential well.

Although a comoving polarizing electron beam may be similar to an
electron cooling beam, the design constraints are different, and
a low beam temperature may be traded for intensity. Clearly, such
an arrangement requires a high-intensity, high-duty-factor,
polarized electron beam, but the rapid developments in this field
of research \cite{card} indicate that this technical limitation may be
overcome.  One clear advantage of a comoving beam is the
possibility to choose the best relative energy between the
protons and the electrons, independent of the energy of the
stored beam. This energy is about 5 MeV (Fig. 1).

It may be possible to produce dense electron targets by plasma
confinement techniques \cite{malm}.  Trapping of electrons in a Penning
trap has been used as a diagnostic tool in the Indiana Cooler. A
realistic extrapolation of the performance of such a device
yields a target thickness of 10$^{12}$ electrons per cm$^2$ .
This is still less than the electron thickness of an internal
hydrogen target. But the task of developing a high-density
electron trap for use in nuclear physics is new and one has to
look to future advances in this area, perhaps borrowing from
techniques developed for plasma confinement in nuclear fusion.
For a trap in which electrons can be {\it accumulated}, the
requirement on the source of polarized electrons is less
demanding. Since the electrons have low velocity in the lab, the
choice of beam energy is critical, trading off the spin transfer
cross section against the lifetime of the beam.

There continues to be great interest in polarized antiproton
beams for nuclear and particle physics research.  Unfortunately,
none of the many methods to polarize antiprotons which have been
suggested \cite{antip} has proven practical.  In $\vec e p$\ scattering
the spin transfer cross section for $\bar p$\ is almost as large
as for $p$, (Fig. 1).  Therefore polarized electron targets may
be useful in polarizing antiprotons.

In this paper we have found a surprisingly large spin-dependent
electromagnetic effect.  It arises from an interference of the
hyperfine amplitude, containing the product of the magnetic
moments of the electron and the proton with the Coulomb
amplitude. The effect is enhanced because of the singularity at
q=0 and because of the large $m_p/m_e$\ mass ratio.  The present
calculation is supported implicitly by the result of the TSR
experiment \cite{rath,meyer}. It would be desirable to test
it further by a similar experiment where the contribution from
the electrons is {\it singled out} by either using a polarized
hydrogen target in atomic spin states that differ only in the
polarization of the electron, by using a mixture of spin states
with polarized electrons, but zero proton polarization, or by
choosing the beam energy such that the polarizing effect from the
pp interaction is small. In any case it should be possible to
carry out such a test using currently existing equipment.

In the present work we use a simple estimate for the effect of
screening. Clearly, at small scattering angles and low energies
this part of the calculation is critical for atomic hydrogen as
well as for pure electron plasma targets. It is important that
the treatment of screening is refined and that the role of
distortions and of external magnetic fields is studied in more
detail.

\acknowledgements
We are indebted to S. Koonin for pointing out the importance of
Coulomb distortions and we gratefully acknowledge useful
discussions with R.E. Pollock. This work has been supported in
part by the US National Science Foundation under grant NSF PHY
93-14783, and by the US Department of Energy under grant
DE-FG0287ER40365.

\begin{figure}
\caption
{Spin transfer cross section $\sigma_{nn}$
in $p\vec e$ scattering
versus beam
kinetic energy $T_{lab}$.  The dotted curves are plane wave Born
approximation (PWBA) results, Eq. (11), which are the same for
protons and antiprotons, while the solid and dashed curves are
approximate distorted wave Born approximation (DWBA)
calculations, Eq. (13), for protons and antiprotons,
respectively.  The upper three curves give the electronic
contribution for an atomic hydrogen target while the lower curves
are for a pure electron target.}
\label{fig1}
\end{figure}

\end{document}